\documentclass[twocolumn]{revtex4}
\usepackage[utf8]{inputenc}
\usepackage{graphicx}

\begin{document}

\title{Analytical analysis of the spin wave dispersion in the cycloidal spin structures under the influence of magneto-electric coupling}

\author{Pavel A. Andreev}
\email{andreevpa@my.msu.ru}
\affiliation{Department of General Physics, Faculty of physics, Lomonosov Moscow State University, Moscow, Russian Federation, 119991.}

\date{\today}

\begin{abstract}
Spin waves and coupling of the spin waves with electromagnetic waves are considered
in the multiferroic materials with the electric dipole moment proportional to the scalar product of spins.
Nature of this interaction is discussed within the spin current model.
Dispersion dependence for the spin waves propagating as the perturbation of the equilibrium state described by spin cycloid is found analytically.
Contribution of the wave vector of the equilibrium cycloid is traced
and it is found that it decreases the contribution of the anisotropy constant,
which can lead to the instability.
Limit regime of the spin waves in systems of collinear spins is described to show the role of the magnetoelectric coupling
for two regimes of the electric dipole moment: it proportional to the scalar product of spins or to the vector product of spins.
The dielectric permeability as the response on the electromagnetic perturbations associated with the magneto-electric coupling for the same equilibrium state is calculated.
\end{abstract}


\maketitle



\section{Introduction}

The electric dipole moment appearance in the nonuniform magnetic structures has been discussed in literature for decades
\cite{Baryakhtar JETP 83}, \cite{Venevtsev 82}, \cite{Venevtsev 79}.
Further systematic study of the magnetoelectricity is made in Refs.
\cite{Jia PRB 06}, \cite{Jia PRB 07}, \cite{Arima JPSC 07}, \cite{Arima PRL 06},
\cite{Sergienko PRL 06}, \cite{Sergienko PRB 06}, \cite{Katsura PRL 07}
\cite{Jia JOpt 19} and reviewed in Refs. \cite{Tokura RPP 14}, \cite{Dong AinP 15}.
Particularly, the possibility of existence of novel type of collective excitations called the electromagnons is suggested
\cite{Smolenskii PU 82} (see also \cite{Katsura PRL 05}),
which are experimentally observed in the low-temperature limit
\cite{Pimenov NP 06}, \cite{Pimenov JP CM 08}, \cite{ShuvaevPimenov EPJB 11}.
It is known that electromagnons (a pair of them) in TbMnO$_{3}$ are observed in the cycloidal magnetic phase
\cite{Aupiais npj QM 18}.

Noncollinear equilibrium spin structures are the essential feature of the multiferroics.
However, both the collinear and the noncollinear structures of spins can form the polarization \cite{Tokura RPP 14}.
Here we are interested in the dynamical properties of spins in the noncollinear structures \cite{Kalva CJP 69}, \cite{JAP 69}
and influence of the magneto-electric coupling associated with the collinear parts of spins on this dynamics.
We are also interested in the dielectric response of the system,
which is shown in the permittivity.
The behavior of permittivity is essential in relation to the detection of the electromagnons.
More complex scenarios regarding electromagnon description are analyzed in recent papers.
In Ref. \cite{Castro PRB 25},
authors describe the dynamics of the electric polarization
(related to the deformation)
in addition to the spin dynamics.
In Ref. \cite{AndreevTrukh EPL 25},
authors consider the collinear equilibrium state of the ferromagnetic samples with the electric dipole moment related to the perpendicular parts of spins,
while main contribution is in the explicit account of external periodic electric field in terms describing magnetoelectric coupling.

The Dzyaloshinskii-Moriya interaction (DMI) is essential for the noncollinear spin structures
and for the formation of the electric dipole moment related to the collinear parts of spins.
Discussion of the collective spin wave excitations under influence of the
Dzyaloshinskii-Moriya interaction
can be found in Refs.
\cite{Moon PRB 13}, \cite{Zakeri PRL 10}.
Simultaneous existence of electromagnetic, spin and acoustic waves and
the hybridization of corresponding dispersion curves
in the ferromagnetic spiral with the account of the Dzyaloshinskii-Moriya interaction
are analysed in Refs. \cite{Bychkov SSP 12}, \cite{Bychkov JMMM 13}.
But the mechanisms of the magneto-electric coupling are not considered in these works.

Analytically simple noncollinear structures are the periodically changing spin density structures described by the trigonometric functions,
such as
\begin{equation}\label{MFMemf}
\textbf{S}_{0}=S_{b}\cos(qx)\textbf{e}_{x}+S_{c}\sin(qx)\textbf{e}_{y}
\end{equation}
for the spin cycloid,
and
\begin{equation}\label{MFMemf}
\textbf{S}_{0}=S_{b}\cos(qz)\textbf{e}_{x}+S_{c}\sin(qz)\textbf{e}_{y}
\end{equation}
for helix.
Here $S_{b}$ and $S_{c}$ are the partial amplitudes of the spin density,
$\textbf{e}_{x}$ and $\textbf{e}_{y}$ are the unit vectors in Cartesian coordinates,
$x$ and $z$ are the coordinates in the corresponding directions,
$q$ is the wave vector of the equilibrium spin structure.

However, more complex structures can exist \cite{Gareeva PRB 13}
\begin{equation}\label{MFMemf sin via sn} \sin\theta=sn(x',\nu)\end{equation}
with $\varphi=const$
(see \cite{Gareeva PRB 13} eq. 25)
and
$$\sin\theta=\frac{\gamma sn(\hat{x},\nu)+1}{sn(\hat{x},\nu)+\gamma}$$
(see \cite{Gareeva PRB 13} eq. 39)
are obtained at the theoretical analysis of phase diagrams in BiFeO$_{3}$-like multiferroics,
where
$sn(x,\nu)$ is the Jacobi elliptical function,
$x'$ and $\hat{x}$ are dimensionless forms of coordinates
(we do not specify it here),
$\nu$ is the elliptical modulus $0 < \nu < 1$ determined from
the minimum of averaged energy (see \cite{Gareeva PRB 13} eq. 26),
$\gamma$ is a combination of parameters of the system,
$\textbf{M}=\textbf{M}_{1}+\textbf{M}_{2}$, $\textbf{L}=\textbf{M}_{1}-\textbf{M}_{2}$,
$\textbf{m}=\textbf{M}/2M_{0}$, $\textbf{l}=\textbf{L}/2M_{0}$,
and 
$\textbf{l}=\sin\theta\cos\varphi\textbf{e}_{x}$
$+\sin\theta\sin\varphi\textbf{e}_{y}$
$+\cos\theta \textbf{e}_{z}$.
Solution (\ref{MFMemf sin via sn}) at fixed $\varphi$ and small parameter $\nu$ appears to be close
to the in-plane cycloid being in plane $\textbf{e}_{\rho}$, $\textbf{e}_{z}$
with $\textbf{e}_{\rho}=\cos\varphi\textbf{e}_{x}+\sin\varphi\textbf{e}_{y}$.
In addition to the direct calculation of the equilibrium states,
soft modes (regimes of frequency goes to zero at nonzero wave vectors) are considered in collinear regimes in order to estimate conditions for the instability of the phase to the transition to the periodic spin orientation equilibrium \cite{Gareeva PRB 13}.

\begin{figure}\includegraphics[width=8cm,angle=0]{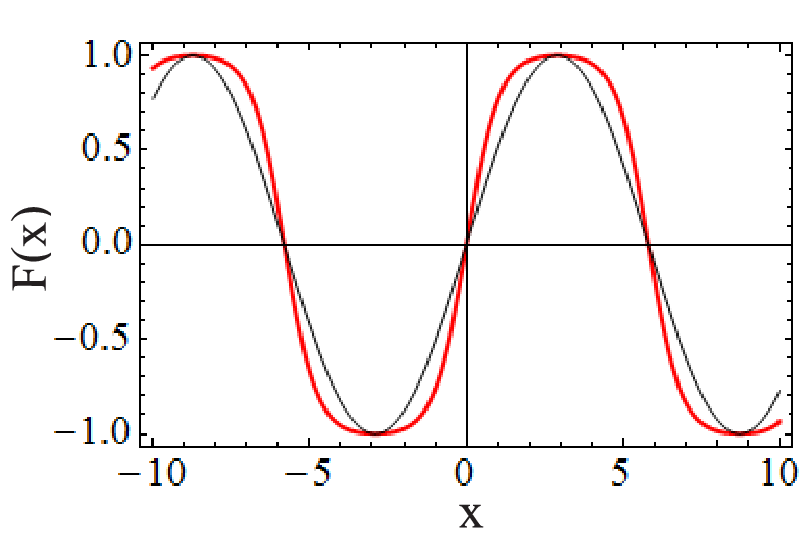}
\caption{\label{MFMextremumP Fig 00}
The figure illustrates comparison of
the Jacobi elliptical function (at $\nu=0.95$)
$F(x)=sn(x,\nu)$ (red thick curve)
and
the trigonometric sin(x) (we choose period to match sin(0.54x) and present it with black thin curve).
At $\nu<0.9$ there is good agreement between sin of corresponding period and $sn(x,\nu)$.
}
\end{figure}

In this paper we are focused on the dispersion dependence of the spin waves existing in noncollinear spin structures.
We consider the magnetization structures described by the trigonometric functions only.
Depending on parameters of the system, the Jacobi elliptical function can be approximated by the trigonometric functions,
while at the large elliptical modulus there is qualitative agreement (see Fig. \ref{MFMextremumP Fig 00}).

The magnon spectrum for the helicoidal state is presented numerically
for the A-type helicoidal magnetic state \cite{Mostovoy PRL 05}.
It shows nonmonotonic behavior with the minimum at nonzero wave vector.
It is specified that the double exchange favors the helicoidal magnetic state with "wave vector" of the equilibrium spiral parallel to one of the cubic axes (A-type helicoidal magnetic state),
while in SrFeO$_{3}$ the helix is parallel to the body diagonal (G-type helicoidal magnetic state) \cite{Mostovoy PRL 05}.

The collective perturbations in the simple spiral
magnets are described in Ref. \cite{Katsura PRL 07},
where
the obtained dispersion dependence for magnons and phonons is based on
the symmetric Heisenberg Hamiltonian and the elasticity evolution,
which are coupled via additional term contributing for the noncollinear spins.
It is interpreted as the magnetoelectric excitations.
Dispersion dependencies for the spin cycloid or helix are considered in Ref. \cite{Fishman PRB 19}.
Some approximations made in the analytical part of this work can be found in appendix
(see eq. A18 and eq A26).

The perturbations in periodic magnetically ordered structures, including the electromagnons
are also considered in Refs. \cite{Aguilar PRL 09}, \cite{Mochizuki PRL 10}, \cite{Goto JJAP 21}, \cite{Kishine PRB 09}.
Particularly, mechanisms of electromagnon excitations are studied for structures appearing in RMnO$_{3}$ materials
(see Ref. \cite{Aguilar PRL 09}).
The contribution of the higher harmonic components of the spiral spin order is discussed in Ref. \cite{Mochizuki PRL 10}.
The dispersion dependence of the spin wave  is discussed in Refs. \cite{Goto JJAP 21} and \cite{Kishine PRB 09}
(see eq 5 in Ref. \cite{Goto JJAP 21} for the 1D helimagnet).

In Ref. \cite{Cherkasskii 24}
authors study the collective excitations of the conical spin spiral equilibrium state,
and obtain the dispersion dependence of the spin waves.
Authors consider both the traceless two-site anisotropy tensor
(it has five independent elements)
and the single-site anisotropy tensor
in addition to the Heisenberg Hamiltonian, DMI, and the Zeeman energy.
They also include the torque proportional to the second time derivative
(it is interpreted as the inertial relaxation time)
in addition to the Gilbert damping.

Below (see equation (\ref{MFMemf disp eq NOSa NOa})) we found the dispersion dependence for the spin wave perturbations of the equilibrium state of spin cycloid
for the easy-plane regime (no damping limit is presented here)
\begin{equation}\label{MFMemf}
\omega^{2}=Ak^2 S_{0}^2 [\mid\kappa\mid-Aq^2 +Ak^2 \mp q \tilde{\delta}]
, \end{equation}
where
$\omega$ is the frequency of the plane wave excitation,
$k$ is the wave vector module,
$A>0$ is the ferromagnetic exchange constant
$S_{0}=\mid S_{b}\mid$ is the amplitude of the equilibrium spiral,
$q$ is the "wave vector" of the equilibrium spiral,
$\kappa$ is the anisotropy interaction constant,
$\mp$ is the difference in signs of projections of the spin density in spiral $S_{c}=\pm S_{b}$,
$\tilde{\delta}$ is the modified shift of the ligand ion entering the Dzyaloshinskii constant.
If we neglect the Dzyaloshinskii-Moriya interaction $\sim \tilde{\delta}$,
it shows the decrease of the phase velocity $\sqrt{A(\mid\kappa\mid-Aq^2)}S_{0}$ in compare with the dispersion dependence of the spin wave perturbations of the equilibrium state of collinear spins
$\sqrt{A\mid\kappa\mid}S_{0}$.

Simple combination of the interactions
(and the consideration of the ferromagnetic material as the most simple magnetically ordered material)
leads to the possibility of the spin wave in the easy-axis regime
(see equation (\ref{MFMemf disp dep with YiX NOa k NOd}) below)
\begin{equation}\label{MFMemf}
\omega^{2}=\frac{1}{2}A S_{b}^{2}(q^{2} -k^{2})
\biggl(\kappa  - A k^2 \mp 2A q  k \biggr)\end{equation}
(no damping limit).

Overall, the small amplitude collective excitations (including spin waves) are the fundamental characteristics of physical systems.
Hence, they are obtained for the ferromagnetic materials
(locally collinear samples)
showing periodic modulation of spin density.
In addition, it is demonstrated that the cycloidal periodic space structure forms no analog of Brillouin zones usual form the wave excitations in periodic structures.

Another item being in focus of this work is the study of the magnetoelectric coupling appearing due to the parallel parts of the spins.
Our analysis leads to the analytical form of the corresponding permittivity with some numerical illustration.

This paper is organized as follows.
In Sec. II mechanisms of electric polarization of the spin origin are discussed, the role of the spin current model is point out. Main focus is made for the regime of collinear spin.
In Sec. III contribution of the magnetoelectric coupling in the macroscopic Landau--Lifshitz--Gilbert equation.
In Sec. IV dispersion dependence for the cycloid equilibrium magnetization order
is derived analytically in the easy-plane samples.
In Sec. V spin waves for the cycloid equilibrium for the easy-axis samples are calculated.
In Sec. VI contribution of the electromagnetic waves is included in the dynamic of spin waves and permittivity is calculated in this case.
In Sec. VII a brief summary of obtained results is presented.

\section{Macroscopic spin polarization of the spin origin}

There are three main mechanisms of the electric polarization appearance in the multiferroic materials
\cite{Tokura RPP 14} (see Fig. 2 on page 3).
The first of these mechanisms corresponds to the electric dipole moment
$\hat{\textbf{d}}_{ij}$ proportional to the scalar product of spins
$\hat{\textbf{s}}_{i}$ and $\hat{\textbf{s}}_{j}$:
\begin{equation}\label{MFMemf edm operator simm}
\hat{\textbf{d}}_{ij}= \mbox{\boldmath $\Pi$}_{ij} (\hat{\textbf{s}}_{i}\cdot\hat{\textbf{s}}_{j}), \end{equation}
where the vector constant $\mbox{\boldmath $\Pi$}_{ij}$ is introduced.
The spin-current model is developed for this regime is Ref. \cite{AndreevTrukh PS 24},
where it is also demonstrated
that the electric dipole moment (\ref{MFMemf edm operator simm}) appears due to the
Dzyaloshinskii-Moriya interaction.
The vector constant $\mbox{\boldmath $\Pi$}_{ij}$
is associated with the element of the
Dzyaloshinskii constant
$\textbf{D}_{ij}=\beta(r_{ij})[\textbf{r}_{ij}\times\mbox{\boldmath $\delta$}]$
\cite{Khomskii JETP 21},
where
$\textbf{r}_{ij}$ is the relative position of two magnetic ions,
$\mbox{\boldmath $\delta$}$ is the ligand shift vector,
i.e. the shift of the nonmagnetic ion from the center of the line connection two magnetic ions,
$\beta(r_{ij})$ is an analog of the exchange integral for the Dzyaloshinskii-Moriya interaction,
which is the scalar isotropic function of the relative distance.
So, the following relation is found in Ref. \cite{AndreevTrukh PS 24}
\begin{equation}\label{MFMemf Pi via beta}\mbox{\boldmath $\Pi$}_{ij}=r_{ij}^2\beta(r_{ij})\mbox{\boldmath $\delta$}\end{equation}
(this is simple of two forms presented in Ref. \cite{AndreevTrukh PS 24}).
Originally spin current model is suggested in Ref. \cite{Katsura PRL 05} for another regime of the polarization formation.
A recent discussion of the spin current model developed in Ref. \cite{Katsura PRL 05} can be found in review \cite{Burns AdvM 20}.

In this Sec. we focus on the derivation of the contribution of the electric dipole moment (\ref{MFMemf edm operator simm}) in
the mean-field Landau--Lifshitz--Gilbert equation.
Therefore, we need the definition of the spin density
\begin{equation}\label{MFMemf S def} \textbf{S}(\textbf{r},t)=
\int \Psi_{S}^{\dagger}(R,t)\sum_{i}\delta(\textbf{r}-\textbf{r}_{i})
(\hat{\textbf{s}}_{i}\Psi(R,t))_{S}dR. \end{equation}
Equation (\ref{MFMemf S def}) contains the following notations:
the LLG equation is formulated for the spin density vector field $\textbf{S}(\textbf{r},t)$,
which is the vector function of space coordinates and time,
It is defined via the many-particle wave function (wave spinor) of all magnetic ions
$\Psi_{S}(R,t)=\Psi_{S}(\textbf{r}_{1}, ..., \textbf{r}_{i}, ..., \textbf{r}_{N},t)$
so $R=\{\textbf{r}_{1}, ..., \textbf{r}_{i}, ..., \textbf{r}_{N}\}$ is the vector in $3N$ dimensional space
showing the aggregate of coordinates of all magnetic ions,
$N$ is the number of the magnetic ions in the system,
$S=\{s_{1}, ..., s_{i}, ..., s_{N}\}$ is the aggregate of spin indexes of all magnetic ions,
$\Psi_{S}^{\dagger}$ is the Hermitian conjugated wave spinor,
$\int ... dR$ is the integral in $3N$ dimensional space,
with $dR=d^{3}r_{1}...d^{3}r_{N}$,
$\hat{\textbf{s}}_{i}$ is the spin operator of $i$-th magnetic ion,
and
$\delta(\textbf{r}-\textbf{r}_{i})$ is the delta-function.

Our derivation, including definition (\ref{MFMemf S def}),
is based on the quantum hydrodynamic method suggested in Ref. \cite{MaksimovTMP 2001}.
Application of this method to the derivation of the mean-field Landau--Lifshitz--Gilbert equation can be found in Ref. \cite{Andreev 2025 Vestn}.
Some additional methodological details are describe in Ref. \cite{Andreev LP 21 fermions}.

Similarly to the spin density (\ref{MFMemf S def}),
we introduce the definition of the electric polarization
\begin{equation}\label{MFMemf P def} \textbf{P}(\textbf{r},t)=
\int \Psi_{S}^{\dagger}(R,t)\sum_{i}\delta(\textbf{r}-\textbf{r}_{i})
(\hat{\textbf{d}}_{i}\Psi(R,t))_{S}dR, \end{equation}
with the corresponding replacement of the operator under quantum average.
In equation (\ref{MFMemf P def}) we use the electric dipole moment operator associated with single ion $i$:
\begin{equation}\label{MFMemf edm operator Mod}
\hat{\textbf{d}}_{i}=\sum_{j\neq i}
\mbox{\boldmath $\Pi$}_{ij}(r_{ij}) (\hat{\textbf{s}}_{i}\cdot\hat{\textbf{s}}_{j}). \end{equation}
Polarization (\ref{MFMemf P def}) can be approximately calculated to find its representation via the spin density
\begin{equation}\label{MFMemf P appr Symm}
\textbf{P}(\textbf{r},t)= \mbox{\boldmath $\delta$}[ c_{0}(\textbf{S}\cdot\textbf{S})+c_{2}(\textbf{S}\cdot\triangle\textbf{S})]
, \end{equation}
where the nonuniform contribution proportional to the space derivatives of the spin density $\triangle\textbf{S}$ is included as well.
This contribution includes both change of the spins in space and the variation of the density of the medium.
Constants $c_{0}$ and $c_{2}$ appear as moments of function :
$c_{0}\mbox{\boldmath $\delta$}=\int \mbox{\boldmath $\Pi$}_{ij}(r_{ij})d^{3}r_{ij}$
and
$c_{2}\mbox{\boldmath $\delta$}=(1/6)\int r_{ij}^{2}\mbox{\boldmath $\Pi$}_{ij}(r_{ij})d^{3}r_{ij}$
(volume integrals on the relative distance $r_{ij}$)
and reduces to the moments of function $\beta(r)$ (\ref{MFMemf Pi via beta}) being a part of the Dzyaloshinskii constant
$\textbf{D}_{ij}=\beta(r_{ij})[\textbf{r}_{ij}\times\mbox{\boldmath $\delta$}]$
(see eq 17 in Ref. \cite{AndreevTrukh PS 24}).

Since the spin density and other macroscopic functions are defined via the single particle spin operator,
let us present the major parameters involved.
We start this description with the commutator
\begin{equation}\label{MFMemf commutator of spins}
[\hat{s}_{i}^{\alpha},\hat{s}_{j}^{\beta}]=\imath\hbar\delta_{ij}\varepsilon^{\alpha\beta\gamma} \hat{s}_{i}^{\gamma}, \end{equation}
where
$\alpha$, $\beta$, $\gamma$ are the tensor indexes,
so each of them is equal to $x$, $y$, $z$,
summation on the repeating Greek symbol is assumed,
$\imath$ is the imaginary unit $\imath^{2}=-1$,
$\delta_{ij}$ is the three-dimensional Kronecker symbol,
$\varepsilon^{\alpha\beta\gamma}$ is the three-dimensional Levi-Civita symbol.

The balance of forces giving the stability of the lattice
includes the balance of the electric dipole-dipole interaction and the spin-orbit interaction.
The balance of these forces can exist at arbitrary inhomogenuity of the electric field.
This leads to the following relation between polarization
(the electric dipole moment density) $\textbf{P}$
and the spin current density $J^{\alpha\beta}$ \cite{AndreevTrukh PS 24}:
\begin{equation}\label{MFMemf}P^{\mu}
=\frac{ \gamma}{c}\varepsilon^{\mu\alpha\beta}J^{\alpha\beta}.
\end{equation}
This equation represent the spin current model of the appearance of the polarization of the spin origin
\cite{Sergienko PRL 06}, \cite{Sergienko PRB 06},
\cite{Tokura RPP 14}, \cite{Dong AinP 15}.

Further, following Ref. \cite{AndreevTrukh PS 24},
we can substitute the (magnon) spin current related to the Heisenberg Hamiltonian
and obtain the polarization
\begin{equation}\label{MFMemf P HH fin} P^{\mu}_{HH}
=\frac{ \gamma }{c}\varepsilon^{\mu\alpha\beta}J^{\alpha\beta}_{HH}
=\frac{ \gamma }{c}g_{u}(S^{\beta}\partial_{\beta} S^{\mu}-S^{\mu}\partial_{\beta} S^{\beta}),\end{equation}
where $g_{u}$ is the parameter depending on the exchange integral entering the Heisenberg Hamiltonian
(see Ref. \cite{AndreevTrukh PS 24}).
This equation corresponds to the well-known result presented in
Refs. \cite{Sparavigna PRB 94}, \cite{Mostovoy PRL 06}.

Otherwise, we can substitute the (magnon) spin current related to the Dzyaloshinskii-Moriya interaction
\begin{equation}\label{MFMemf P DM fin} P^{\mu}_{DM}
=\frac{ \gamma }{c}\varepsilon^{\mu\alpha\beta}J^{\alpha\beta}_{DM}
=\frac{ \gamma }{c} \frac{1}{3}g_{2(\beta)}
\biggl[(\mbox{\boldmath $\delta$}\cdot\textbf{S})S^{\mu}-\frac{1}{2}\delta^{\mu}\textbf{S}^{2}
\biggr],\end{equation}
where $g_{2(\beta)}$ and $\mbox{\boldmath $\delta$}$ depend on the elements of the vector Dzyaloshinskii constant \cite{AndreevTrukh PS 24}.

The spin current in its nature can be related to two major classes.
The first major class, it is the transition of particles caring the spin in space.
The second major class, it is the effective spin current related to the transfer of spin by the spin waves.
The first of these major classes can be caused by three mechanisms
\cite{Andreev PTEP 19} (see eq. 9).
The first mechanism is related to the migration of the spins with the macroscopic velocity
(it can reveal itself in the fluids or gases, or in the electron gas).
The second mechanism is associated with the displacement of the spins at the thermal motion of particles.
The third mechanism is based on the quantum nature of particles.
It is similar to the quantum Bohm potential.
The quantum spin current has the following approximate form
\begin{equation}\label{MFMemf spin current Bohm}
J^{\alpha\beta}_{Bohm}=-\frac{1}{m}
\varepsilon^{\alpha\mu\nu}S^{\mu}\partial^{\beta}\biggl(\frac{S^{\nu}}{n}\biggr). \end{equation}
In accordance with the spin current model of polarization,
it leads to the following polarization
\cite{AndreevTrukh JETP 24}
$$P^{\mu}_{Bohm}
=\frac{ \gamma }{c}
\varepsilon^{\mu\alpha\beta}J^{\alpha\beta}_{Bohm}$$
\begin{equation}\label{MFMemf P Bohm fin}
=\frac{\gamma }{mc}
\biggl[S^{\mu}\partial^{\beta}\biggl(\frac{S^{\beta}}{n}\biggr)-S^{\beta}\partial^{\beta}\biggl(\frac{S^{\mu}}{n}\biggr)\biggr].\end{equation}
If the variations of the density (including the phonons) are not generated in system
this spin current is analogous to the spin current caused by the
Heisenberg Hamiltonian,
otherwise it gives additional contribution related to the density non-homogeneity or the phonon dynamics.

To avoid any confusion,
we need to mention a recent mini-review
\cite{Mostovoy npj 24},
where near equations 8 and 9,
mechanisms of the polarization formation are discussed.
Presented in Ref. \cite{Mostovoy npj 24} understanding of physical mechanisms differs from the presented above.

Our concept follows from the generalized spin current model \cite{AndreevTrukh JETP 24}, \cite{AndreevTrukh PS 24}.
It is related to the balance of forces of the electric dipoles with the local electric field
and the spin-orbit interaction of the spin currents with the local electric field.
Appearance of the electric dipole moment (small deformations) ensures the stability of the system.
So, the polarization forms to balance the spin-orbit interaction,
which causes the polarization in the first place.
The spin-orbit interaction force depends on the form of the real or effective spin current.
It gives different interplays between electric dipole moment and the spin configurations (forming the effective spin currents).
Therefore, it gives different forms of the electric polarization of spin origin described above.

Basically, we describe a self-consistent process of formation of configurations of spins and the deformation of ion crystals (polarization),
including some reorganization of the local electric field.

In described picture, we approximately assume
that the exchange integrals changing accordingly to the reconstruction of system
(at the phase transition or further change of temperature).
However, the exchange integrals are considered as fix (unknown) functions for the fixed temperature.
These functions give the interaction constants like $c_{0}$ and $c_{2}$,
which can be found from experiments.
Hence, no direct relation of the exchange integrals and corresponding interaction constants on the electric field is assumed.

If we apply the electric field it would lead to the reorganization of charged particles.
It obviously induces some polarization of the system.
It would change the exchange integrals as well.
It can affect the spin configuration as well.
This polarization is the addition to the spin caused polarization and additional polarization may depend on the spin configuration.
It is possible that the model described in Ref. \cite{Mostovoy npj 24} deals with this mechanism,
since explicit dependence of the exchange integrals (and similar coefficients) on the electric field is assumed.

\section{Model: Macroscopic Landau--Lifshitz--Gilbert equation}

Let us present the mean-field Landau--Lifshitz--Gilbert equation,
which contains the contribution of the magneto-electric coupling,
applied in this paper
$$\partial_{t}\textbf{S}=
A[\textbf{S}\times\triangle\textbf{S}]
+\kappa [\textbf{S}\times S_{z}\textbf{e}_{z}] $$
$$+\frac{1}{3}g_{(\beta)}\biggl((\textbf{S}\cdot[\mbox{\boldmath $\delta$}\times\nabla])\textbf{S}
-\frac{1}{2}[\mbox{\boldmath $\delta$}\times\nabla]S^{2}\biggr) $$
$$+2 c_{2}[\textbf{S}\times
((\mbox{\boldmath $\delta$}\cdot \textbf{E})\triangle \textbf{S}
+(\mbox{\boldmath $\delta$}\cdot(\partial^{\nu} \textbf{E}))\cdot\partial^{\nu} \textbf{S})]$$
\begin{equation}\label{MFMemf s evolution MAIN TEXT}
+a[\textbf{S}\times\partial_{t}\textbf{S}]. \end{equation}
The evolution of spin density happens due to the interparticle interaction manifesting itself
as the spin-torque placed on the right-hand side of the spin evolution equation.
The first term describes the isotropic exchange interaction.
The second term is the contribution of the anisotropy energy related to the exchange interaction
and being the consequence of the anisotropy mainly created by the elastic forces and therefore affecting
overlapping of the wave functions in different directions.
The third term is the Dzyaloshinskii-Moriya interaction corresponding to the vector form of the Dzyaloshinskii constant considered
in the Keffer form,
where
the displacement of the ligand $\mbox{\boldmath $\delta$}$ is explicitly included
(see eq. 17 in Ref. \cite{Dong AinP 15}).
This term can be obtained using the quantum hydrodynamic method with no explicit account of the lattice structure
\cite{AndreevTrukh PS 24},
while the account of structure of multiferroic bismuth ferrite and the spin cycloid is made in Ref.
\cite{ZvezdinPyatakov EPL 12} during the derivation of corresponding macroscopic free energy,
which is also known from 1982
\cite{Sosnowska JP C 82}.
There is the form of Dzyaloshinskii-Moriya interaction,
where the Dzyaloshinskii constant is proportional to the interparticle distance $\textbf{D}_{ij}\sim \textbf{r}_{ij}$
(see for instance eq.
14 in Ref. \cite{Fishman PRB 19}).
It gives additional spin torque,
but it does not lead to any polarization (at least in the spin current model),
so we do not consider it here.
The fourth term is the contribution of the magneto-electric coupling for the parallel spins considered above.
It corresponds to polarization (\ref{MFMemf P appr Symm}).
This contribution can be found at the calculation of the spin density evolution by the quantum hydrodynamic method \cite{AndreevTrukh PS 24} and \cite{AndreevTrukh EPJ B 24} with additional term in Hamiltonian describing action of the electric field on the electric dipole moment
(\ref{MFMemf edm operator Mod}).
Or it can be found using the variational principle with
the additional term in the energy density of the system
$\Delta \mathcal{E}=-\textbf{P}\cdot\textbf{E}$,
where $\textbf{P}$ is the polarization or the electric dipole moment density (\ref{MFMemf P appr Symm}),
and $\textbf{E}$ is the electric field.
The last term is the Gilbert damping with negative constant $a<0$.

We can compare the spin torque existing in the system of collinear spins (presented in equation (\ref{MFMemf s evolution MAIN TEXT}))
with its analog form the noncollinear spins
\cite{Risinggard SR 16}, \cite{Andreev 2025 05}
$$\textbf{T}=-\sigma
\biggl[ [\textbf{E}\times \nabla] S^{2}
-2(\textbf{S}\cdot[\textbf{E}\times\nabla]) \textbf{S}
$$
\begin{equation}\label{MFMemf T for non coll spins}-S^{2}(\nabla\times\textbf{E})
+\textbf{S}(\textbf{S}\cdot [\nabla\times\textbf{E}])
\biggr],
\end{equation}
which corresponds to polarization \cite{Sparavigna PRB 94}, \cite{Mostovoy PRL 06}
\begin{equation}\label{MFMemf P def expanded} \textbf{P}(\textbf{r},t)=
\sigma
[\textbf{S}(\nabla\cdot \textbf{S})-(\textbf{S}\cdot\nabla)\textbf{S}]. \end{equation}
We see the difference in the vector structures, number of derivatives,
and the presence of the characteristic direction $\mbox{\boldmath $\delta$}$ for the collinear regime (\ref{MFMemf s evolution MAIN TEXT}).

\subsection{Contribution of the electric polarization in the dispersion dependence of spin waves}

This paper is mainly focused on the spin waves relatively nonuniform periodic equilibrium structures.
But, in this subsection we consider the spin wave relatively uniform collinear equilibrium.
It specifies the contribution of the magnetoelectric coupling in the spin waves in simple cases for the further comparison with the nonuniform equilibrium.
Moreover, we do not include the perturbations of the electric field,
but we consider the constant uniform external electric field $\textbf{E}_{0}$.
Here and below we assume that the anisotropy axis is directed parallel $z$-axis.

The structure of the spin torque caused by the magnetoelectric coupling with the "perpendicular" spins (\ref{MFMemf T for non coll spins})
is similar to the structure of the DMI caused torque
(for the constant electric field and at the replacement $2\sigma \textbf{E}\rightarrow (1/3)g_{(\beta)}\mbox{\boldmath $\delta$}$).
This similarity can follow from the fact that the electric field induces the polarization
(additional relative shift of magnetic and nonmagnetic ions with the opposite sign of charges,
additional $\mbox{\boldmath $\delta$}$ in other words)
and this additional polarization induces the DMI-like interaction.
It also gives similar contribution to the dispersion dependence of the spin waves.

The structure of the spin torque caused by the magnetoelectric coupling with the "parallel" spins
is similar to the structure of the symmetric Heisenberg Hamiltonian caused torque
(at the replacement $2c_{2}(\mbox{\boldmath $\delta$}\cdot \textbf{E}_{0})\rightarrow A$).
For the interpretation we consider the polarization induced by the electric field
$\mbox{\boldmath $\delta$}_{eff}\sim \textbf{E}_{0}$.
Hence, presented interaction is a contact interaction of the electric dipoles modulated by the spin density.
Obviously, two described in this paragraph interactions give similar contribution to the dispersion dependence of the spin waves.

\subsubsection{Polarization related to the collinear parts of spins}

In this subsubsection, we consider the spin waves for the collinear equilibrium.
We use LLG equation (\ref{MFMemf s evolution MAIN TEXT})
with the magnetoelectric coupling presented with the term proportional to $c_{2}$.
Polarization in this case has maximum in the equilibrium state,
while the dynamics of spin leads to the decrease of polarization.

\emph{Easy-axis: $\textbf{S}_{0}=S_{0}\textbf{e}_{z}$ and $\kappa>0$.}

First we present the dispersion dependence for the easy-axis regime
$$\omega=\frac{1}{1-\imath a S_{0}}\times$$
\begin{equation}\label{MFMemf }
\times\biggl(\kappa+[A+2c_{2}(\mbox{\boldmath $\delta$}\cdot \textbf{E}_{0})]k^2
+\frac{1}{3}g_{(\beta)}(k_{x}\delta_{y}-k_{y}\delta_{x})\biggr)S_{0}.
\end{equation}
Here, the MEC gives the electric field dependent modification of the exchange constant
$A+2c_{2}(\mbox{\boldmath $\delta$}\cdot \textbf{E}_{0})$,
while the DMI $\sim g_{(\beta)}$ shows the well-known nonreciprocity.

\emph{Easy-plane: $\textbf{S}_{0}=S_{0}\textbf{e}_{x}$ and $\kappa<0$.}

The easy-plane regime shows similar modification of the dispersion dependence via the modification of the exchange constant:
$$[\omega-(1/3)g_{(\beta)}(k_{y}\delta_{z}-k_{z}\delta_{y})S_{0}]^2$$
$$=\biggl((A+2c_{2}(\mbox{\boldmath $\delta$}\cdot \textbf{E}_{0}))k^2+\imath\omega a\biggr)\times$$
\begin{equation}\label{MFMemf DE CR EP}
\times\biggl(\mid\kappa\mid +(A+2c_{2}(\mbox{\boldmath $\delta$}\cdot \textbf{E}_{0}))k^2+\imath\omega a\biggr),
\end{equation}
where we present the dispersion equation for the nonzero damping.
If we neglect the damping
$a=0$
we can get explicit form of the dispersion dependence
$$\omega=\biggl[\sqrt{(A+2c_{2}(\mbox{\boldmath $\delta$}\cdot \textbf{E}_{0}))[\mid\kappa\mid +(A+2c_{2}(\mbox{\boldmath $\delta$}\cdot \textbf{E}_{0}))k^2]}k$$
\begin{equation}\label{MFMemf DD CR EP}
+\frac{1}{3}g_{(\beta)}(k_{y}\delta_{z}-k_{z}\delta_{y})\biggr]S_{0}.\end{equation}
The first term is modified with the magnetoelectric coupling,
while the last term presents the DMI with no contribution of the polarization.

\subsubsection{Polarization related to the noncollinear parts of spins}

In this subsubsection we use LLG equation (\ref{MFMemf s evolution MAIN TEXT}),
but the magnetoelectric coupling is given by equation (\ref{MFMemf T for non coll spins}), instead of the term proportional to $c_{2}$.

In the collinear equilibrium,
we have no polarization for this mechanism of the polarization formation,
but the polarization appears at the spin dynamics.

\emph{Easy-axis: $\textbf{S}_{0}=S_{0}\textbf{e}_{z}$ and $\kappa>0$.}

General structure of the dispersion dependence of the spin waves for the easy-axis regime is conserved
$$\omega=\frac{1}{1-\imath a S_{0}}\biggl(\kappa+Ak^2$$
\begin{equation}\label{MFMemf }
+\frac{1}{3}g_{(\beta)}(k_{x}\delta_{y}-k_{y}\delta_{x})+2\sigma (k_{x}E_{0y}-k_{y}E_{0x})\biggr)S_{0},
\end{equation}
but the electric field shows the formation of the electric dipole moment $d_{extra,i}\sim \sigma E_{0i}$,
with $i=x,y$,
which works similarly to the ligand shift $\delta_{i}$ existing in the system (and in the DMI Hamiltonian) before the electric field application.

\emph{Easy-plane: $\textbf{S}_{0}=S_{0}\textbf{e}_{x}$ and $\kappa<0$.}

To avoid large expressions similar to equation (\ref{MFMemf DE CR EP})
we consider the zero damping only $a=0$
and demonstrate the dispersion dependence
$$\omega=\biggl[\sqrt{A[\mid\kappa\mid +Ak^2]}k$$
\begin{equation}\label{MFMemf }
+\frac{1}{3}g_{(\beta)}(k_{y}\delta_{z}-k_{z}\delta_{y})+2\sigma(k_{y}E_{0z}-k_{z}E_{0y})\biggr]S_{0}.\end{equation}
Similarly to the easy-axis configuration (\ref{MFMemf DD CR EP}) we see that
the contribution of the magnetoelectric coupling $\sim \sigma$ is analogous to the contribution of the DMI.

In order to distinguish the contribution of magnetoelectric coupling from the DMI
(or from the symmetric exchange in the previous subsubsection)
we can trace the change of the dispersion dependence at the change of the applied electric field
(neglecting the possible modification of the exchange integrals).
Particularly, if the ligand shift is oriented parallel to one coordinate axis
($\delta_{z}\neq0$, $\delta_{y}=0$, for instance),
we can apply the electric field in the second direction $E_{0y}\neq0$.
This is a principal illustration for the formal assumption of cubic symmetry.
Real samples would require special analysis.

\section{Dispersion dependence for the cycloid equilibrium magnetization order}

Existence of the electromagnons in the cycloidal magnetic phase for TbMnO$_{3}$ \cite{Aupiais npj QM 18} shows
the necessity of the analysis of the collective excitations in this state.

\subsection{Cycloid order for the easy-plane samples}

\emph{Equilibrium}:
Let us start this subsection with the analysis of the possible equilibrium condition
within the chosen anzatz of the cycloid spin order
\begin{equation}\label{MFMemf S0 eq NO Sa}
\textbf{S}_{0}=S_{b}\cos(qx)\textbf{e}_{x}+S_{c}\sin(qx)\textbf{e}_{y}
. \end{equation}
This cycloid is placed in the plane which is perpendicular to the anisotropy axis.
Next, we need to find relations between parameters $S_{c}$, $S_{b}$, and $q$.
We consider equilibrium regime of the LLG equation with the zero external electric field.
Hence, we put $\partial_{t}\textbf{S}_{0}=0$ and $\textbf{E}_{0}=0$, and then find
$$A[\textbf{S}_{0}\times\triangle\textbf{S}_{0}]
+\kappa [\textbf{S}_{0}\times S_{0z}\textbf{e}_{z}] $$
\begin{equation}\label{MFMemf s evolution eq 1}
+\frac{1}{3}g_{(\beta)}\biggl((\textbf{S}_{0}\cdot[\mbox{\boldmath $\delta$}\times\nabla])\textbf{S}_{0}
-\frac{1}{2}[\mbox{\boldmath $\delta$}\times\nabla]S_{0}^{2}\biggr)=0. \end{equation}
We include $\triangle\textbf{S}_{0}=-q^{2}\textbf{S}_{0}$ so the first term is equal to zero.
Next, we have $S_{0z}=0$ so the second term is equal to zero.
The third term is equal to zero
$(\textbf{S}_{0}\cdot[\mbox{\boldmath $\delta$}\times\nabla])$
$\sim(\textbf{S}_{0}\cdot[\mbox{\boldmath $\delta$}\times\textbf{e}_{x}])=0$
if we satisfy one of the following conditions
1) $\mbox{\boldmath $\delta$}\parallel\textbf{e}_{x}$
or
2) $\mbox{\boldmath $\delta$}\parallel\textbf{e}_{y}$ so
$\textbf{S}_{0}\perp[\mbox{\boldmath $\delta$}\times\textbf{e}_{x}]$.
If
$\mbox{\boldmath $\delta$}\parallel\textbf{e}_{x}$
the last term is equal to zero as well,
but
if $\mbox{\boldmath $\delta$}\parallel\textbf{e}_{y}$
the last term goes to zero at
$S_{c}^{2}=S_{b}^{2}$.

\emph{Perturbations}:

Let us consider the small amplitude perturbations for the found equilibrium.
Here we consider $\mbox{\boldmath $\delta$}\parallel\textbf{e}_{y}$
or the Dzyaloshinskii-Moriya interaction gives the zero contribution in the calculations.
We decompose the spin density as the equilibrium part described above $\textbf{S}_{0}$
and perturbations $\delta \textbf{S}\ll \textbf{S}_{0}$:
$\textbf{S}= \textbf{S}_{0}+ \delta \textbf{S}$.
This form is substituted in Landau--Lifshitz--Gilbert equation (\ref{MFMemf s evolution MAIN TEXT}).
Nonlinear terms on the perturbations are neglected.

Some details are considered in Appendix A,
it gives equation for
$\delta S_{z}$ with constant coefficients.
So, it can be solved using the
Fourier transform
$$\omega^{2}(1+a^2 S_{0}^2)+\imath\omega a S_{0}^2[\kappa+A(q^2 -2k^2) \pm q \tilde{\delta}]$$
\begin{equation}\label{MFMemf disp eq NOSa}
=-Ak^2 S_{0}^2[\kappa+A(q^2-k^2) \pm q \tilde{\delta}]
. \end{equation}
If we put $a=0$
we find
\begin{equation}\label{MFMemf disp eq NOSa NOa}
\omega^{2}=Ak^2 S_{0}^2 [\mid\kappa\mid-Aq^2 +Ak^2 \mp q \tilde{\delta}]
, \end{equation}
with
$\kappa<0$ corresponding to the easy-plane regime.
It shows us the linear dispersion dependence similar to the easy-plane regime,
but shifted towards smaller frequencies and phase velocities by term $Aq^2$.
The DMI also gives the contribution $\mp q \tilde{\delta}$ independent of the wave vector of the perturbation $k$.
Similar shift appears in the imaginary part of frequency if we include nonzero damping $a$.
Let us also point out that sign of $\tilde{\delta}$ can be positive or negative
depending on sign of function $\beta$ entering the microscopic DMI.

Dispersion dependence (\ref{MFMemf disp eq NOSa NOa}) shows instability of the cycloid spin order
for the relatively small anisotropy constant $\mid\kappa\mid$ and large enough $Aq^2$.

In contrast to the collinear order with spins parallel to Ox direction, where $\delta S_{x}=0$,
we see that all three components of the spin density evolve in the spin wave perturbation.
However, final equation appears for the nontrivial dynamics of the spin density parallel to the anisotropy axis $\delta S_{z}$.

\section{Dispersion dependence for the cycloid equilibrium magnetization order for the easy-axis samples}

We found the plane wave perturbations for the spin density parallel to the anisotropy axis $\delta S_{z}$.
Other projection can demonstrate more complex behavior for the considered mode (found for the easy-axis regime).
Let us consider the possibility of another mode,
where dynamics is related to the plane wave perturbations of the spin density projections
perpendicular to the anisotropy axis $\delta S_{x}$ and $\delta S_{y}$.
We continue the analysis of the set of differential equations (\ref{MFMemf dSx inPl}), (\ref{MFMemf dSy inPl}), and (\ref{MFMemf dSz inPl}) with
changing coefficients,
where coefficients depend on coordinate $x$.
Hence, we can consider the harmonic oscillations of the spin density in time $\delta\textbf{S}=\textbf{S}(x)e^{-\imath\omega t}$
with unspecified dependence on coordinate $x$.

For this regime we consider some technical steps in Appendix B.
Equations (\ref{MFMemf eq for mathcal X}) and (\ref{MFMemf eq for mathcal Y}) show some symmetry of coefficients.
It points out that there is a combination of $\delta S_{x}$ and $\delta S_{y}$
which is a solution of these equations.

Hence, the solution of equation (\ref{MFMemf eq diff for YiX}) gives us the
dispersion equation
$$2\omega^{2}-S_{b}^{2}QM-S_{b}^{2}Q\tilde{\delta}k$$
\begin{equation}\label{MFMemf disp eq with YiX a k d}
+AS_{b}^{2}[q^2 M+k^2 M+ 2qk (\pm1)M +q^{2}k\tilde{\delta} +k^3 \tilde{\delta}
+2q (\pm1)\tilde{\delta}k^2], \end{equation}
where
\begin{equation}\label{MFMemf M}
M= q^{2}A-\imath\omega a \pm q\tilde{\delta}-Ak^{2}.
\end{equation}

First let us consider the limit of neglecting
the DMI
$$2\omega^{2}-S_{b}^{2}Q'M'$$
\begin{equation}\label{MFMemf disp eq with YiX a k NOd}
+AS_{b}^{2}M'[q^2+k^2 + 2qk (\pm1)], \end{equation}
where
\begin{equation}\label{MFMemf M'}
M'= q^{2}A-\imath\omega a -Ak^{2},
\end{equation}
and
\begin{equation}\label{MFMemf Q'}
Q'= q^{2}A+\kappa-\imath\omega a .
\end{equation}

Next, let us consider the regime of the zero damping $a=0$:
$$2\omega^{2}-S_{b}^{2}Q''M''$$
\begin{equation}\label{MFMemf disp eq with YiX NOa k NOd}
+AS_{b}^{2}M''[q^2+k^2 + 2qk (\pm1)], \end{equation}
where
\begin{equation}\label{MFMemf M''}
M''= A(q^{2} -k^{2}),
\end{equation}
and
\begin{equation}\label{MFMemf Q''}
Q''= q^{2}A+\kappa.
\end{equation}

Equation (\ref{MFMemf disp eq with YiX NOa k NOd}) allows to understand main properties of the spin waves in the considered regime.
It shows the frequency of the zero wave vector limit $k=0$
\begin{equation}\label{MFMemf disp dep with YiX NOa NOk NOd}
\omega^{2}=A\kappa S_{b}^{2}q^2/2. \end{equation}
Hence, this wave exist for the easy-axis magnetic materials.
Moreover, the existence of the nonzero frequency in $k=0$ limit is related to the periodic (cycloid) equilibrium spin order with the period $q$.

Let us present the dispersion dependence following from equation
(\ref{MFMemf disp eq with YiX NOa k NOd})
\begin{equation}\label{MFMemf disp dep with YiX NOa k NOd}
\omega^{2}=\frac{1}{2}A S_{b}^{2}(q^{2} -k^{2})
\biggl(\kappa  - A k^2 \mp 2A q  k \biggr). \end{equation}

\begin{figure}\includegraphics[width=8cm,angle=0]{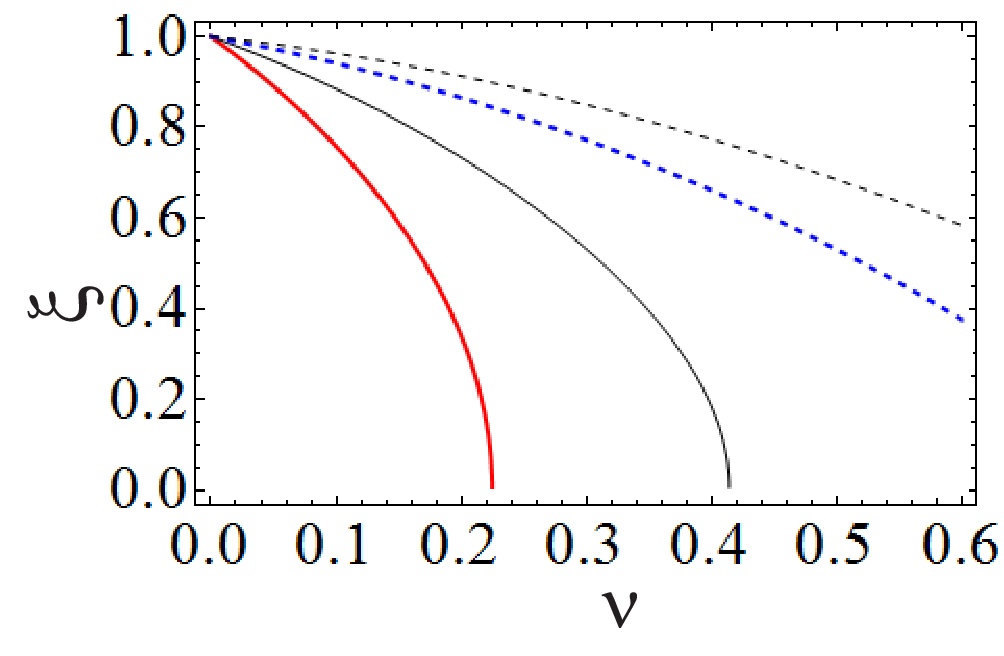}
\caption{\label{MFMextremumP Fig 01} The figure shows the dispersion dependence of spin wave in cycloidal structure in the easy-axis regime
for the equal signs of the cycloid amplitudes $S_c=S_b$ in accordance with equation
(\ref{MFMemf disp dep with YiX NOa k NOd dimless}).
Here $\xi=\omega/\omega_{0}$ is the dimensionless frequency,
and $\nu=k/q$ is the dimensionless wave vector.
Parameter $r=Aq^{2}/\kappa$ chosen to be equal to $r_1=2$ (the lower thick continuous red line),
$r_2=1$ (the second from below thin continuous black line),
$r_3=0.5$ (the second from above thick dashed blue line),
$r_4=0.3$ (the upper thin dashed black line).}
\end{figure}

\begin{figure}\includegraphics[width=8cm,angle=0]{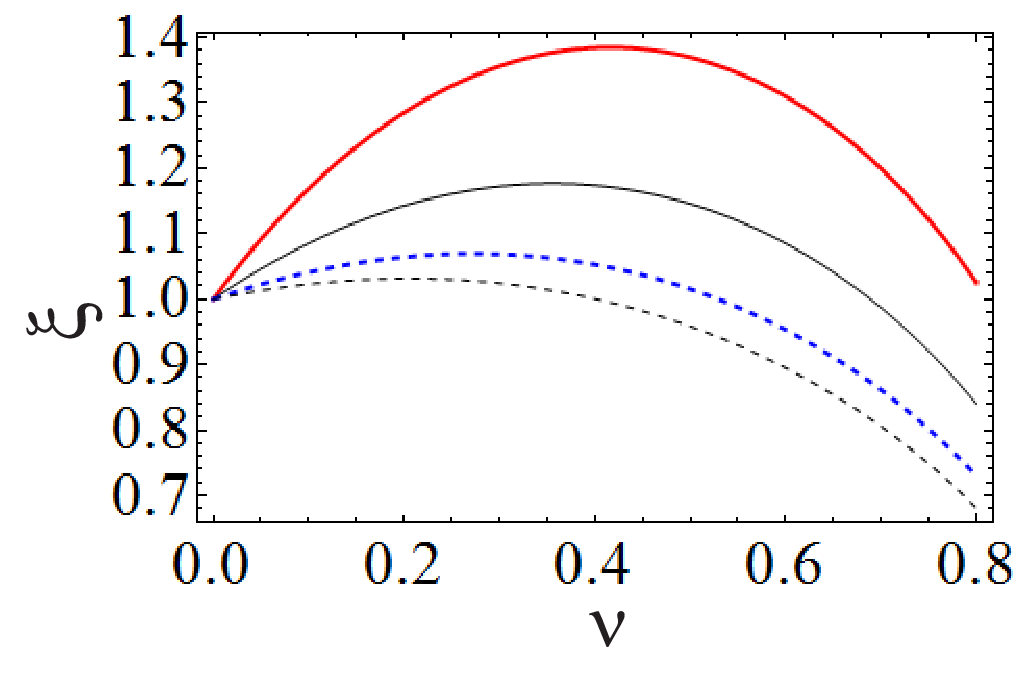}
\caption{\label{MFMextremumP Fig 02} The figure shows the dispersion dependence of spin wave in cycloidal structure in the easy-axis regime
for the different signs of the cycloid amplitudes $S_{c}=-S_{b}$ in accordance with equation
(\ref{MFMemf disp dep with YiX NOa k NOd dimless}).
Parameter $r=Aq^{2}/\kappa$ chosen to be equal to $r_{1}=2$ (the upper thick continuous red line),
$r_{2}=1$ (the second from above thin continuous black line),
$r_{3}=0.5$ (the second from below thick dashed blue line),
$r_{4}=0.3$ (the lower thin dashed black line). }
\end{figure}

In order to analyze two dispersion dependencies presented within equation (\ref{MFMemf disp dep with YiX NOa k NOd})
we consider its dimensionless form and show the result of numerical analysis as well
\ref{MFMextremumP Fig 01} and \ref{MFMextremumP Fig 02}.
Here we use the following notations
for the frequency $\xi\equiv \omega/\omega_{0}$,
with $\omega_{0}=\sqrt{A\kappa S_{b}^{2}q^2/2}$,
and the wave vector module $\nu=k/q$.
Hence equation (\ref{MFMemf disp dep with YiX NOa k NOd}) can be represented in the following form
\begin{equation}\label{MFMemf disp dep with YiX NOa k NOd dimless}
\xi^{2}=(1-\nu^2)
(1 - r \nu^2  \mp r \nu), \end{equation}
where $r=Aq^{2}/\kappa$.
Hence, we have single dimensionless parameter $r=Aq^{2}/\kappa$ affecting the dispersion dependence.
Parameters $\omega_{0}$ and $q$ appears to be most natural units for the reverse time and length.
Hence, we expect that the dimensionless parameter $r=Aq^{2}/\kappa$ can be chosen close to $1$, at the numerical analysis.

Consider the upper sign in equation (\ref{MFMemf disp dep with YiX NOa k NOd dimless})
corresponding to $S_{c}=S_{b}$.
It shows that all terms lead to the decrease of the frequency at the increase of the wave vector.
However, we see stable spectrum up to the wave vector equal to the wave vector $q$ of the equilibrium cycloid for the large coefficient of the anisotropy.
While, the decrease of the anisotropy in comparison with the exchange constant leads to the decrease of the critical wave vector.
It is demonstrated in
Fig.
\ref{MFMextremumP Fig 01},
where
the variation of the dispersion dependence as the function of parameter $r=Aq^{2}/\kappa$.

Next, consider the lower sign in equation (\ref{MFMemf disp dep with YiX NOa k NOd dimless})
corresponding to $S_{c}=-S_{b}$.
There is the competition of terms depending on the wave vector.
In the small wave vector regime we have domination of the lower term with the positive sign,
while other (negative) terms dominate at larger wave vectors.
Hence, there is area of the decline of the dispersion dependence down to the zero value at $k=q$.
Fig.
\ref{MFMextremumP Fig 02}
shows that increase of parameter $r=Aq^{2}/\kappa$ leads
to the increase of the area of increase of the dispersion dependence and $k_{max}$ and $\omega_{max}$ also increase.

\section{Dielectric permeability for the easy-plane multiferroics}

In this section we consider
dielectric permeability and the dispersion equation for the spin and the electromagnetic waves.

Next we consider the perturbations of the electric field along with the perturbations of the spin density
(in the linear regime on the small amplitude perturbations).
We use equation (\ref{MFMemf s evolution MAIN TEXT}),
where terms containing the electric field give nonzero contribution and represent the magneto-electric effect.
Let us specify the linearized form of the terms containing the perturbations of the electric field
$\partial_{t}\textbf{S}_{\Pi,lin}$$
=c_{2}\{(\mbox{\boldmath $\delta$}\cdot\delta \textbf{E})[\textbf{S}_{0},\triangle\textbf{S}_{0}]$$
+2(\mbox{\boldmath $\delta$}\cdot \partial^{\gamma}\delta \textbf{E})[\textbf{S}_{0},\partial^{\gamma}\textbf{S}_{0}] \}$,
with
$[\textbf{S}_{0},\triangle\textbf{S}_{0}]=0$
and
$[\textbf{S}_{0},\partial^{\gamma}\textbf{S}_{0}]=\delta^{x\gamma}\textbf{e}_{z} qS_{b}S_{c}$.
So, the following term appears
$\partial_{t}\textbf{S}_{\Pi,lin}=2c_{2}qS_{b}S_{c}\cdot\delta\cdot\delta^{z\alpha}\partial_{x}\delta E_{y}$.

Consider the perturbations of the polarization
$$\delta \textbf{P}= \mbox{\boldmath $\delta$}
[2c_{0}(\textbf{S}_{0}\cdot\delta \textbf{S}) $$
\begin{equation}\label{MFMemf delta P via delta S bold}
+c_{2}(\textbf{S}_{0}\cdot\triangle\delta \textbf{S}) +c_{2}(\delta \textbf{S}\cdot\triangle\textbf{S}_{0})],
\end{equation}
where the first term and the last term are equal to zero.
We find
$(\textbf{S}_{0}\cdot\triangle\delta \textbf{S})$
$=(S_{0x}\partial_{x}^{2}\delta S_{x}+S_{0y}\partial_{x}^{2}\delta S_{y})$
$=\imath\cdot 2qS_{b}S_{c}\hat{L}\delta S_{z}/\omega$,
where $\hat{L}=\hat{M}+\kappa$
(\ref{MFMemf M hat}).

Hence, the final expression for the polarization perturbation is
\begin{equation}\label{MFMemf delta P via delta Sz}
\delta \textbf{P}= \imath\cdot \mbox{\boldmath $\delta$}
\frac{2q c_{2}S_{b}S_{c}}{\omega}(q^{2}A-\imath\omega a \pm q\tilde{\delta}+A\partial_{x}^{2}+\kappa)\delta S_{z},
\end{equation}
with
$\mbox{\boldmath $\delta$}=\delta\cdot \textbf{e}_{y}$.

\begin{figure}\includegraphics[width=8cm,angle=0]{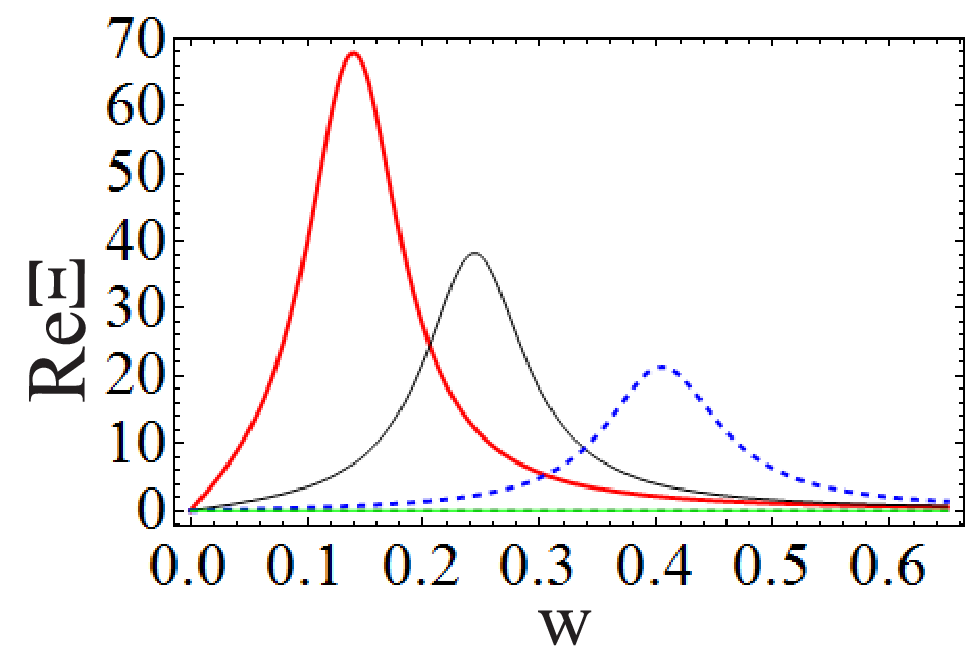}
\caption{\label{MFMextremumP Fig 03} The figure shows
the dimensionless form
for the real part of the
dielectric permeability
$Re\Xi=\kappa^{yy}_{R}/(4q^2 S_{b}^{2}S_{c}^{2}c_{2}^{2}\delta^{2}k \mid\kappa\mid)$
(\ref{MFMemf kappa yy Re dimless})
as the function of the dimensionless frequency $\textrm{w}=\omega/(\mid\kappa\mid S_{b})$.
The dimensionless wave vector of the spin cycloid is equal $\tilde{q}=\sqrt{A/\mid\kappa\mid}q=0.2$.
The wave vector (dimensionless) of the perturbation is chosen as the parameter
$\tilde{k}=\sqrt{A/\mid\kappa\mid}k$:
$\tilde{k}=0.15$ for the red continuous line,
$\tilde{k}=0.25$ for the black thin continuous line,
and
$\tilde{k}=0.4$ for the blue dashed line.
The damping constant is chosen to be $aS_{b}=-0.1$.
The Dzyaloshinskii-Moriya interaction contribution in $\mathcal{M}$ is dropped
in this estimation as the correction to the Heisenberg exchange interaction.}
\end{figure}

\begin{figure}\includegraphics[width=8cm,angle=0]{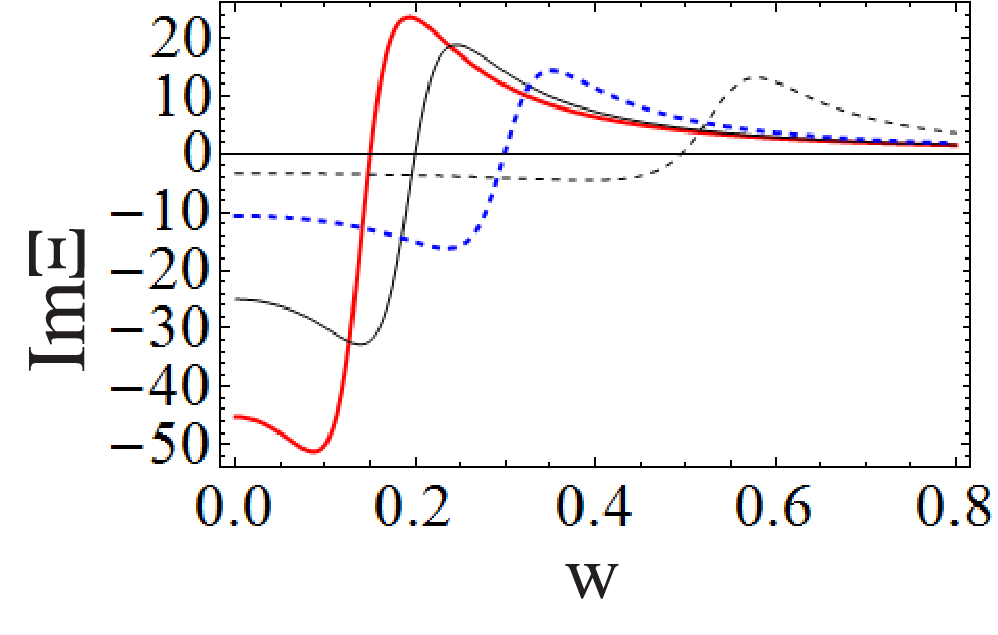}
\caption{\label{MFMextremumP Fig 04} The figure shows
the dimensionless form
for the imaginary part of the dielectric permeability
(\ref{MFMemf kappa yy Im dimless})
as the function of the dimensionless frequency $\textrm{w}=\omega/(\mid\kappa\mid S_{b})$.
The dimensionless wave vector of the spin cycloid is equal $\tilde{q}=\sqrt{A/\mid\kappa\mid}q=0.2$.
The wave vector (dimensionless) of the perturbation is chosen as the parameter
$\tilde{k}=\sqrt{A/\mid\kappa\mid}k$:
$\tilde{k}=0.15$ for the red continuous  line,
$\tilde{k}=0.2$ for the the black thin continuous  line,
$\tilde{k}=0.3$ for the blue dashed  line,
and
$\tilde{k}=0.5$ for black thin dashed line.
The damping constant is chosen to be $aS_{b}=-0.1$.
The Dzyaloshinskii-Moriya interaction contribution in $\mathcal{M}$ is dropped
in this estimation as the correction to the Heisenberg exchange interaction.}
\end{figure}

\subsection{Maxwell's equations}

In order to consider the influence of the polarization on the electromagnetic waves we need to include
Maxwell's equations,
which lead to the following wave equation for the electric field
\begin{equation}\label{MFMemf WE from ME}
\partial_{x}^2\delta E_{y}+\frac{\omega^2}{c^2}\delta E_{y}
-\frac{4\pi\gamma\imath\omega}{c}\partial_{x}\delta S_{z}
+4\pi\frac{\omega^2}{c^2}\delta P_{y}
=0. \end{equation}
We include it here to consider the presence or absence of the periodic coefficients in this equation.

Let us present generalization of wave equation (\ref{MFMemf omega delta Sz})
appearing at the account of the electric field perturbations
$$-\omega^{2}\delta S_{z}+\imath\omega\cdot2c_{2}qS_{b}S_{c}\cdot\delta\cdot\partial_{x}\delta E_{y}$$
\begin{equation}\label{MFMemf}
=-(A\partial_{x}^2-\imath\omega a)S_{b}^{2}(Aq^2+\kappa-\imath\omega a +A\partial_{x}^2 \pm q \tilde{\delta})\delta S_{z}.
\end{equation}

We find a coupled set of equations for $\delta S_{z}$ and $\delta E_{y}$ with constant coefficients.
Hence, we can solve it using the
Fourier transform
in order to find the algebraic relations between $\delta S_{z}$ and $\delta E_{y}$:
\begin{equation}\label{MFMemf}
\delta S_{z}= \frac{-2\omega k c_{2} qS_{b}S_{c}\cdot\delta\cdot\delta E_{y}}{\omega^{2}+S_{0}^2 (Ak^2+\imath\omega a)[-\imath\omega a+\kappa+Aq^2-Ak^2 \pm q \tilde{\delta}]}
,
\end{equation}
where the denominator
goes to zero if the frequency satisfy dispersion equation for the spin waves
(\ref{MFMemf omega delta Sz})
or (\ref{MFMemf disp eq NOSa}).

\subsection{Dielectric permeability}

In the considered regime,
we obtain the single element for the dielectric permeability tensor
$\delta P^{\alpha}=$
$(\varepsilon^{\alpha\beta}-\delta^{\alpha\beta})\delta E^{\beta}/4\pi$
$=\kappa^{\alpha\beta}\delta E^{\beta}$,
which is $\kappa^{yy}$:
\begin{equation}\label{MFMemf}
\kappa^{yy}=\frac{4\imath\cdot q^{2} k c_{2}^{2} S_{b}^{2}S_{c}^{2}\cdot\delta^{2}(\mid\kappa\mid-Aq^2+Ak^2 \mp q \tilde{\delta}+\imath\omega a)}{\omega^{2}+S_{0}^2 (Ak^2+\imath\omega a)[-\imath\omega a+\kappa+Aq^2-Ak^2 \pm q \tilde{\delta}]}. \end{equation}

The separation of the dielectric permeability on the real and imaginary parts is made in the following form
$$\kappa^{yy}=\kappa^{yy}_{R}+\imath \kappa^{yy}_{Im}$$
Hence, we obtain the following form for the imaginary part
near the frequency of spin wave and in the long-wavelength limit
$$\kappa^{yy}_{Im}(\omega=\omega_{R},k\rightarrow0)
=-\frac{4q^{2}c_{2}^{2}\delta^{2}}{\omega_{R}^{2}}\frac{Ak^3}{1+a^2S_{0}^{2}} $$
\begin{equation}\label{MFMemf}
=-\frac{4q^{2}c_{2}^{2}\delta^{2}k}{\mid\kappa\mid-Aq^2}.\end{equation}

Contribution of $\nabla\times \textbf{S}$ can give additional contribution in the extended dielectric permeability
$\delta P_{y,eff}=\gamma\frac{kc}{\omega}\delta S_{z}$.
It can be considered as the contribution of the magnetic permeability,
but it defines the properties on the refractive index anyway.

Let us present the explicit forms for the real and imaginary parts of the dielectric permeability
$$\kappa^{yy}_{R}=-4q^2 S_{b}^{2}S_{c}^{2}c_{2}^{2}\delta^{2}k\omega a\times$$
\begin{equation}\label{MFMemf kappa yy Re}\times
\frac{\omega^2(1+a^2 S_{b}^{2})+\mathcal{M}^{2}}{[\omega^2(1+a^2 S_{b}^{2})- Ak^2 S_{b}\mathcal{M}]^2 +\omega^{2}a^2 S_{b}^{2}[Ak^2 S_{b}+\mathcal{M}]^2} , \end{equation}
where
$\mathcal{M}=(\mid\kappa\mid -Aq^2 +A k^2 \mp q \tilde{\delta})S_{b}$,
$a=-\mid a\mid$
and
$$\kappa^{yy}_{Im}=4q^2 S_{b}S_{c}^{2}c_{2}^{2}\delta^{2}k\times$$
\begin{equation}\label{MFMemf kappa yy Im}
\times
\frac{\omega^2 \mathcal{M}-Ak^2 S_{b}(\mathcal{M}^{2}+\omega^{2}a^2 S_{b}^{2}) }{[\omega^2(1+a^2 S_{b}^{2})- Ak^2 S_{b}\mathcal{M}]^2 +\omega^{2}a^2 S_{b}^{2}[Ak^2 S_{b}+\mathcal{M}]^2}. \end{equation}

Next, we introduce the dimensionless parameters and show
the dimensionless form
for the real part of the dielectric permeability
$$Re\Xi\equiv\frac{\kappa^{yy}_{R}}{4q^2 S_{b}^{2}S_{c}^{2}c_{2}^{2}\delta^{2}k \mid\kappa\mid}$$
\begin{equation}\label{MFMemf kappa yy Re dimless}
=\frac{w \mid a\mid S_{b}(w^2(1+a^2 S_{b}^{2})+m^{2})}{[w^2(1+a^2 S_{b}^{2})-m \tilde{k}^2 ]^2 +w^{2}a^2 S_{b}^{2}[\tilde{k}^2+m]^2}
, \end{equation}
where
$\tilde{k} = \sqrt{A /\mid\kappa\mid}k$,
$w=\omega/(\mid\kappa\mid S_{b})$,
$m=\mathcal{M}/(\mid\kappa\mid S_{b}).$

Behavior of $Re\Xi$ given by equation (\ref{MFMemf kappa yy Re dimless}) is illustrated in
Fig. \ref{MFMextremumP Fig 03},
where the contribution of the cycloidal spiral
(appearing mainly via the Heisenberg exchange interaction)
is chosen to be comparable with the contribution of the anisotropy energy.
Overwise, the noncollinear equilibrium order of spins can be neglected and considered approximately as the collinear structure.
The real part of the dielectric permeability shows single peak related to the eigenfrequency of the spin wave.
The figure is given for several values of the perturbation wave vector $k$ near the equilibrium cycloidal spiral wave vector $q$.
The increase of the wave vector $k$ obviously leads to the increase of frequency (\ref{MFMemf disp eq NOSa NOa}),
but it is accompanied with the decrease of the dielectric response.
No specific behavior can be detected near $k=q$.

We also present
the dimensionless form
for the imaginary part of the dielectric permeability
$$Im\Xi\equiv\frac{\kappa^{yy}_{Im}}{4q^2 S_{b}^{2}S_{c}^{2}c_{2}^{2}\delta^{2}k \mid\kappa\mid}$$
\begin{equation}\label{MFMemf kappa yy Im dimless}
=\frac{m w^2-\tilde{k}^{2}(m^{2}+w^2 a^2 S_{b}^{2})}{[w^2(1+a^2 S_{b}^{2})-m \tilde{k}^2 ]^2 +w^{2}a^2 S_{b}^{2}[\tilde{k}^2+m]^2}
. \end{equation}

Behavior of $Im\Xi$ given by equation (\ref{MFMemf kappa yy Im dimless}) is illustrated in
Fig. \ref{MFMextremumP Fig 04}
in the regime similar to previous figure for the real part.
Here we see s-like curve with changing sign,
while the dispersion dependence (\ref{MFMemf disp eq NOSa}) is stable and shows the standard damping.
Larger part of the curve tends to be in the negative area (positive area) for relatively small (large) $k$.
The middle point is about $\tilde{k}\approx 2\tilde{q}=0.4$.

Figures for $Re\Xi$ and $Im\Xi$ look like they are rearranged.
Polarization perturbations are expressed via perturbations of the spin density with real coefficient
(\ref{MFMemf delta P via delta S bold}),
but nonzero contributions appears via $\delta S_{x}$ and $\delta S_{y}$.
Further representation via $\delta S_{z}$ leads to appearance of the imaginary unit $\imath$
(\ref{MFMemf delta P via delta Sz}).
It rearranges the contribution of the real and imaginary parts of the denominator of $\delta S_{z}(\delta E_{y})$
and explain demonstrated behavior.

\section{Conclusion}

The contribution of the magneto-electric coupling existing between parallel parts of spins in the dynamic of the spiral spin structures has been considered.
Corresponding spin torque has been derived.
It consists of two terms.
One of them shows similarity to the exchange term with the coefficient depending on the electric field
$2 c_{2}(\mbox{\boldmath $\delta$}\cdot \textbf{E})[\textbf{S}\times\triangle \textbf{S}]$.
The second part contains the first space derivative of the spin density,
while the coefficient depends on the derivative of the electric field
$2 c_{2}(\mbox{\boldmath $\delta$}\cdot(\partial^{\nu} \textbf{E}))[\textbf{S}\times\partial^{\nu} \textbf{S})]$.
Resent development of the spin current model of the polarization appearance in the multiferroics of spin origin has been reviewed.
It shows
that two main mechanisms of the polarization formation can be explained within the spin current model
and demonstrates
that the contribution of collinear spins in the polarization is related to the Dzyaloshinskii-Moriya interaction (in a combination with the spin-orbit interaction)
and
the contribution of noncollinear spins in the polarization is caused by the symmetric Heisenberg exchange interaction.

Spin waves in the cycloid equilibrium spin structures have been analyzed using the macroscopic
Landau-Lifshitz-Gilbert equation.
Corresponding dispersion equations have been obtained analytically with no account of the magneto-electric coupling
and under influence of the magneto-electric coupling.
Cycloidal equilibrium keeps same structure of $\omega(k)$ as in the collinear limit for the easy-plane samples.
But the cycloidal equilibrium modifies some coefficients in $\omega(k)$.
These modifications are proportional to the wave vector $q$ of the equilibrium cycloid
and tend to decrease the contribution of the anisotropy constant $\kappa$.
It can lead to the instability $\omega^{2}<0$ for small enough anisotropy constant
if $\mid\kappa\mid -Aq^{2}<0$,
where $A$ is the exchange constant.
Moreover, the second spin wave solution has been found,
which exist in the cycloidal equilibrium and has no analogues in the collinear equilibrium.
Its frequency defined by the wave vector of the equilibrium cycloid at the zero wave vector of perturbations.
Particularly, the dielectric permeability of the electromagnetic waves in these systems has been derived and discussed.

\section{DATA AVAILABILITY}

Data sharing is not applicable to this article as no new data were
created or analyzed in this study, which is a purely theoretical one.

\section{Acknowledgements}

The work is supported by the Russian Science Foundation under the
grant
No. 25-22-00064.

\appendix

\section{Cycloid order for the easy-plane samples: some calculations}

Vector form of the linearized Landau--Lifshitz--Gilbert equation
$$\partial_{t}\delta \textbf{S}=A[\textbf{S}_{0},\triangle\delta \textbf{S}]
+A[\delta \textbf{S},\triangle \textbf{S}_{0}]
-S_{0}^{\beta}[\tilde{\mbox{\boldmath $\delta$}}\times\textbf{e}_{x}]\partial_{x}\delta S^{\beta}$$
$$+(\delta \textbf{S}\cdot[\tilde{\mbox{\boldmath $\delta$}}\times\textbf{e}_{x}])\partial_{x}\textbf{S}_{0}
-\delta S^{\beta}[\tilde{\mbox{\boldmath $\delta$}}\times\textbf{e}_{x}]\partial_{x}S_{0}^{\beta}
$$
\begin{equation}\label{MFMemf linearized Landau--Lifshitz--Gilbert equation}
+\kappa[ \textbf{S}_{0},\delta S_{z}\textbf{e}_{z}]
+a[\textbf{S}_{0},\partial_{t}\delta \textbf{S}]
, \end{equation}
with
$\triangle \textbf{S}_{0}=-q^2\textbf{S}_{0}$,
$\tilde{\mbox{\boldmath $\delta$}}= (1/3)g_{(\beta)}\mbox{\boldmath $\delta$}$,
and
$\mbox{\boldmath $\delta$}=\delta\cdot \textbf{e}_{y}$.

The first line in equation (\ref{MFMemf linearized Landau--Lifshitz--Gilbert equation})
shows two terms
$A[\textbf{S}_{0},\triangle\delta \textbf{S}]
+A[\delta \textbf{S},\triangle \textbf{S}_{0}]$
appearing from the exchange term
$A[\textbf{S},\triangle \textbf{S}]$.
On the qualitative level we expect $\triangle\delta \textbf{S}=-k^{2}\delta \textbf{S}$,
where $k$ is the wave vector module.
It gives
$-A\{ k^2[\textbf{S}_{0},\delta \textbf{S}]
+q^{2}[\delta \textbf{S}, \textbf{S}_{0}]\}$
$=A(q^{2}-k^2)[\textbf{S}_{0},\delta \textbf{S}]$,
so these terms gives terms with opposite signs in the dispersion equation.

Consider the projections of the linearized Landau--Lifshitz--Gilbert equation.
We also include $\delta\textbf{S}=\textbf{S}(x)e^{-\imath\omega t}$ since coefficients of the differential equations depend on coordinate $x$ only.
The $x-$ and $y-$ projections show
that they can be expressed via the perturbations of the $z-$ projection:
\begin{equation}\label{MFMemf dSx inPl}
\imath\omega\delta S_{x}=-S_{c}\sin(qx)[\kappa+Aq^2+A\triangle \pm q \tilde{\delta}-\imath\omega a]\delta S_{z}
, \end{equation}
and
\begin{equation}\label{MFMemf dSy inPl}
\imath\omega\delta S_{y}=S_{b}\cos(qx)[\kappa+Aq^2+A\triangle \pm q \tilde{\delta}-\imath\omega a]\delta S_{z}
. \end{equation}
It shows the phase shift of $\delta S_{x}$ from $\delta S_{y}$ on $\pi/2$.

The $z$-projection
can be completely expressed via $x-$ and $y-$ projections
$$-\imath\omega\delta S_{z}=S_{b}\cos(qx)\tilde{\delta}\partial_{x}\delta S_{x} +S_{c}\sin(qx) \tilde{\delta}\partial_{x}\delta S_{y}$$
$$+S_{b}\cos(qx)[-\imath\omega a+Aq^2+A\triangle \pm q \tilde{\delta}]\delta S_{y}$$
\begin{equation}\label{MFMemf dSz inPl}
-S_{c}\sin(qx)[-\imath\omega a+Aq^2+A\triangle \pm q \tilde{\delta}]\delta S_{x}
, \end{equation}
where $S_{c}=\pm S_{b}$.

We can put expressions for $\delta S_{x}$ (\ref{MFMemf dSx inPl}) and $\delta S_{y}$ (\ref{MFMemf dSy inPl})
in equation for $\delta S_{z}$ (\ref{MFMemf dSz inPl}).
So, we exclude all functions except $\delta S_{z}$.
Moreover, periodic coefficients combine to each other in order to give the constant coefficients
$$\omega^{2}\delta S_{z}=
-\imath\omega a S_{0}^2[-\imath\omega a+\kappa+Aq^2+A\triangle \pm q \tilde{\delta}]\delta S_{z}$$
\begin{equation}\label{MFMemf omega delta Sz}
+S_{0}^2 A[-\imath\omega a+\kappa+Aq^2+A\triangle \pm q \tilde{\delta}]\triangle\delta S_{z}
. \end{equation}
This is the result of straightforward substitution with no approximation.

\section{Cycloid order for the easy-axis samples: some calculations}

After substitution of equation (\ref{MFMemf dSz inPl}) in equation (\ref{MFMemf dSx inPl})
we obtain the following equation
$$\mathcal{X}\equiv\omega^{2}\delta S_{x}
+S_{0x}  S_{0y} Q \hat{M} \delta S_{y} -QS_{0y}^{2} \hat{M} \delta S_{x}$$
$$+S_{0x}  S_{0y} Q \tilde{\delta}\partial_{x} \delta S_{x} +QS_{0y}^{2} \tilde{\delta}\partial_{x} \delta S_{y}
+S_{0y} A\biggl[
-q^{2}S_{0x}\hat{M} \delta S_{y}$$
$$+S_{0x}\partial_{x}^{2}(\hat{M} \delta S_{y})-2q S_{0y}(\pm1)\partial_{x}(\hat{M} \delta S_{y})$$
$$+q^{2}S_{0y}\hat{M} \delta S_{x}-S_{0y}\partial_{x}^{2}(\hat{M} \delta S_{x})-2q S_{0x}(\pm1)\partial_{x}(\hat{M} \delta S_{x})$$
$$-q^{2}S_{0x}\tilde{\delta}\partial_{x} \delta S_{x} +S_{0x}\tilde{\delta}\partial_{x}^{3} \delta S_{x}
-2qS_{0y}\tilde{\delta}(\pm1)\partial_{x}^{2}\delta S_{x}$$
\begin{equation}\label{MFMemf eq for mathcal X}
-q^{2}S_{0y}\tilde{\delta}\partial_{x} \delta S_{y} +S_{0y}\tilde{\delta}\partial_{x}^{3} \delta S_{y}
+2qS_{0x}\tilde{\delta}(\pm1)\partial_{x}^{2}\delta S_{y}\biggr]=0, \end{equation}
where
\begin{equation}\label{MFMemf Q}
Q= q^{2}A+\kappa-\imath\omega a \pm q\tilde{\delta},
\end{equation}
and
\begin{equation}\label{MFMemf M hat}
\hat{M}= q^{2}A-\imath\omega a \pm q\tilde{\delta}+A\partial_{x}^{2}.
\end{equation}
After substitution of equation (\ref{MFMemf dSz inPl}) in equation (\ref{MFMemf dSy inPl})
we obtain the following equation
$$\mathcal{Y}\equiv\omega^{2}\delta S_{y}
+S_{0x}  S_{0y} Q \hat{M} \delta S_{x} -QS_{0x}^{2} \hat{M} \delta S_{y}$$
$$-S_{0x}  S_{0y} Q \tilde{\delta}\partial_{x} \delta S_{y} -QS_{0x}^{2} \tilde{\delta}\partial_{x} \delta S_{x}
+S_{0x} A\biggl[
-q^{2}S_{0x}\hat{M} \delta S_{y}$$
$$+S_{0x}\partial_{x}^{2}(\hat{M} \delta S_{y})-2q S_{0y}(\pm1)\partial_{x}(\hat{M} \delta S_{y})$$
$$+q^{2}S_{0y}\hat{M} \delta S_{x}-S_{0y}\partial_{x}^{2}(\hat{M} \delta S_{x})-2q S_{0x}(\pm1)\partial_{x}(\hat{M} \delta S_{x})$$
$$-q^{2}S_{0x}\tilde{\delta}\partial_{x} \delta S_{x} +S_{0x}\tilde{\delta}\partial_{x}^{3} \delta S_{x}
-2qS_{0y}\tilde{\delta}(\pm1)\partial_{x}^{2}\delta S_{x}$$
\begin{equation}\label{MFMemf eq for mathcal Y}
-q^{2}S_{0y}\tilde{\delta}\partial_{x} \delta S_{y} +S_{0y}\tilde{\delta}\partial_{x}^{3} \delta S_{y}
+2qS_{0x}\tilde{\delta}(\pm1)\partial_{x}^{2}\delta S_{y}\biggr]
=0. \end{equation}
Above we introduced notations for the
left-hand sides
of these equations
$\mathcal{X}$ and $\mathcal{Y}$.

We consider
$$\delta S_{y}=\imath\delta S_{x}$$
and also consider
$$\mathcal{Y}+\imath\mathcal{X}=0$$
and obtain:
$$2\omega^{2}\delta S_{x}
+S_{b}^{2}
\biggl(-Q\hat{M}\delta S_{x}+\imath Q \tilde{\delta}\partial_{x} \delta S_{x}
+Aq^2\hat{M}\delta S_{x}$$
$$-A\partial_{x}^2(\hat{M}\delta S_{x})-2q\imath (\pm1)A\partial_{x}(\hat{M}\delta S_{x})
-\imath q^2 A \tilde{\delta}\partial_{x} \delta S_{x}$$
\begin{equation}\label{MFMemf eq diff for YiX}
+\imath\tilde{\delta} A\partial_{x}^3\delta S_{x}
-2q(\pm1)\tilde{\delta}A\partial_{x}^2\delta S_{x}\biggr)=0
. \end{equation}
Described substitution let us to find the differential equation with the constant coefficients.
Consequently,
we use
Fourier transform:
$\delta S_{x}=\mathcal{S}e^{\imath k x}$.

\end{document}